\documentstyle[aps,prd]{revtex}
\def\be{\begin{equation}}
\def\ee{\end{equation}}
\def\ba{\begin{eqnarray}}
\def\ea{\end{eqnarray}}
\begin{document}
\draft 
\title{A Second Poincar\'e Group\footnote{Dedicated
to Abraham Hirsz Zimerman on the occasion of his $70^{\rm th}$ birthday.}}
\author{R. Aldrovandi and J. G. Pereira}
\vskip 0.5cm
\address{Instituto de F\'{\i}sica Te\'orica\\
Universidade Estadual Paulista\\
Rua Pamplona 145\\
01405-900\, S\~ao Paulo \\ 
Brazil}
\maketitle

\begin{abstract}

Solutions of the sourceless Einstein's equation with weak and strong cosmological
constants are discussed by using In\"on\"u-Wigner contractions of the de Sitter groups
and spaces. The more usual case corresponds to a weak cosmological-constant limit, in
which the de Sitter groups are contracted to the Poincar\'e group, and the de Sitter
spaces are reduced to the Minkowski space. In the strong cosmological-constant limit,
however, the de Sitter groups are contracted to another group which has the same
abstract Lie algebra of the Poincar\'e group, and the de Sitter spaces are reduced to a
4-dimensional cone-space of infinite scalar curvature, but vanishing Riemann and
Ricci curvature tensors. In such space, the special conformal transformations act
transitively, and the equivalence between inertial frames is that of special
relativity.

\end{abstract}

\vskip 1.5cm

\section{INTRODUCTION}

The Poin\-ca\-r\'e group $P$ is naturally associated to Minkowski spacetime $M$ as its
group of motions. It contains, in the form of a semi--direct product, the Lorentz group
$L = SO(3,1)$ and the translation group $T$. The latter acts transitively on $M$ and its
manifold is just $M$. Indeed, Minkowski  spacetime is a homogeneous space under $P$,
actually the quotient $M = T = P/L$. If we prefer, the manifold of $P$ is a principal
bundle ${\cal P}(P/L, L)$ with $P/L = M$ as base space and $L$ as the typical
fiber~\cite{kono}.
	
The invariance of $M$ under the transformations of $P$ reflects its uniformity. The
Lorentz subgroup provides an isotropy around a given point of $M$, and the translation
invariance enforces this isotropy around any other point. This is the usual meaning of
``uniformity", in which $T$ is responsible for the equivalence of all points of
spacetime.
	
The concept is actually more general. In all local (or tangential) physics, what happens
is that the laws of Physics are invariant under transformations related to some specific
kind of uniformity of which the above case is but an example (though, quite probably, the
most important one). Uniformity includes homogeneity of space and of time, isotropy of
space and the equivalence of inertial frames. This holds of course for usual
special--relativistic kinematics, but also for Galilean and other non--relativistic
kinematics~\cite{levy}, their difference being grounded in their different ``kinematical
groups". Most of our experiments are local, and presuppose some such kinematics.
	
The complete kinematical group, whatever it may be, will always have a subgroup
accounting for both the isotropy of space (rotation group) and the equivalence of
inertial frames. The remaining transformations will be generically called
``translations", commutative or not. Roughly speaking, the point--set of local spacetime
is the point--set of these ``translations". More precisely, kinematical spacetime is
defined as the quotient space of the whole kinematical group by the subgroup including
rotations and boosts.

Given any solution of Einstein's equation, that is, any acceptable spacetime, it is the
local kinematics which will provide the stage--set for local experiments. Our aim in this
paper will be to prospect new possible relativistic kinematical groups and spacetimes and
reveal an apparently as yet unsuspected case. Our starting point is the well known fact
that the Poincar\'e group can be obtained from the de Sitter group by an appropriate
In\"on\"u--Wigner contraction~\cite{gilmore}. Such contractions have been first
introduced~\cite{inonu} to formalize and generalize the fact that the Galilei group is
obtained from the Lorentz group in the non--relativistic limit $c \rightarrow \infty$.
The general procedure involves always a preliminary choice of convenient coordinates, in
terms of which a certain parameter is made explicit which encapsulates the whole limiting
process --- the complete new kinematics is obtained by taking  that parameter to an
appropriate limit. In the specific case of the contraction of the de Sitter to the
Poincar\'e group, the parameter is the de Sitter pseudo--radius ${\cal R}$, and the limit
is achieved by taking ${\cal R}$ to infinity. The curvature of the de Sitter space,
which is proportional to ${\cal R}^{-2}$, goes consequently to zero in the limit.

Now, the de Sitter space is a solution of Einstein's equation for an empty space with a
cosmological constant $\Lambda = R/4$, where $R$ is the scalar curvature of the de Sitter
space. Therefore, the limit of the de Sitter curvature going to zero is equivalent to the
limit in which the cosmological constant $\Lambda$ goes to zero. In this limit, the de
Sitter groups reduce to the Poincar\'e group, and the de Sitter spaces reduce to Minkowski
space, a sourceless solution of Einstein's equation with a vanishing cosmological
constant. We should mention that, for a small enough cosmological constant, it would be
very difficult to differentiate experimentally between a small $\Lambda$ and a vanishing
$\Lambda$.

A natural question then arises: what happens in the limit of the de Sitter
pseudo--radius ${\cal R}$ going to zero, corresponding to a cosmological constant going
to infinity? In what follows, we will be concerned mainly with this question. We start by
studying the de Sitter groups and spaces. The conformal stereographic coordinates are
introduced in such a way to explicitly exhibit the de Sitter pseudo--radius ${\cal R}$.
This is necessary to apply the contraction procedure. The contraction limit ${\cal R}
\rightarrow \infty$ is then studied, which takes the de Sitter group into the Poincar\'e
group, and the de Sitter space into Minkowski space. Proceeding further, the contraction
limit ${\cal R} \rightarrow 0$ is studied, which is shown to take the de Sitter groups
into a kind of ``conformal" Poincar\'e group --- the ``second Poincar\'e group" of our
title --- and the de Sitter spaces into a $4$-dimensional cone--space. Finally, in the
last section, we discuss a duality relation between the two cases.

\section{THE DE SITTER GROUPS AND SPACES} 
	
Amongst curved spacetimes, only those of constant curvature can lodge the highest number
of Killing vectors. Given the metric signature and the value of the scalar curvature $R$,
these maximally--symmetric spaces are unique~\cite{weinberg}. In consequence, the de
Sitter spaces are the only uniformly curved 4-dimensional metric spacetimes. There are
two kinds of them~\cite{ellis}, one with positive, and another one with negative
curvature. They can be defined as hypersurfaces in the pseudo--Euclidean spaces ${\bf
E}^{4,1}$ and ${\bf E}^{3,2}$, inclusions whose points in Cartesian coordinates $(\xi^A)
= (\xi^0, \xi^1, \xi^2, \xi^3$, $\xi^4)$ satisfy, respectively,
\ba
\eta_{AB} \xi^A \xi^B &=& - (\xi^0)^2 + (\xi^1)^2 + (\xi^2)^2 + (\xi^3)^2 + (\xi^4)^2 =
{\cal R}^2 \; ; \nonumber \\ {} \nonumber \\
\eta_{AB} \xi^A \xi^B &=& - (\xi^0)^2 + (\xi^1)^2 + (\xi^2)^2 + (\xi^3)^2 - (\xi^4)^2 =
- {\cal R}^2 \; . \nonumber
\ea
We use $\eta_{a b}$ (with indices $a, b = 0,1,2,3$) for the Lorentz
metric $\eta = $ diag $(-1$, $1$, $1, 1)$ and the notation $\epsilon = \eta_{44}$ to
put the conditions together as
\be
\eta_{a b} \, \xi^{a} \xi^{b} + \epsilon \left(\xi^4\right)^2 =
\epsilon {\cal R}^2 \; .
\label{dspace}
\ee
 
The de Sitter space $dS(4,1)$, whose metric can be put into the diagonal form $\eta_{AB}$
= $(-1,+1,+1,+1,+1)$, has the pseudo--orthogonal group $SO(4,1)$ as group of motions. The
other, with metric $\eta_{AB}$ = $(-1,+1,+1,+1,-1)$, is frequently called anti--de Sitter
space and is denoted $dS(3,2)$ because its group of motions is $SO(3,2)$. The de Sitter
spaces are both homogeneous spaces:
\[
dS(4,1) = SO(4,1)/ SO(3,1) \quad \mbox{and} \quad dS(3,2) = SO(3,2)/ SO(3,1) \; .
\]
The manifold of each de Sitter group is a bundle with the corresponding de Sitter space
as base space and $L=SO(3,1)$ as fiber. But the kinematical group is no more a product of
groups. If we isolate $L$ and call the remaining transformations ``de Sitter
translations", these do not constitute a subgroup and the product of two of them amounts
to a Lorentz transformation.

The generators of infinitesimal de Sitter transformations are given by
\be
J_{A B} = \eta_{AC} \, \xi^C \, \frac{\partial}{\partial \xi^B} -
\eta_{BC} \, \xi^C \, \frac{\partial}{\partial \xi^A} \; .
\label{dsgene}
\ee
They satisfy the commutation relations
\be
\left[ J_{AB}, J_{CD} \right] = \eta_{BC} \, J_{AD} + \eta_{AD} \, J_{BC}
- \eta_{BD} \, J_{AC} - \eta_{AC} \, J_{BD} \; .
\label{dsal}
\ee

The 4-dimensional stereographic coordinates $x^\mu$  are given
by~\cite{gursey}
\be
\xi^{a} = n(x) \, \delta^{a}{}_{\mu} \, x^\mu \equiv h^{a}{}_{\mu} \, x^\mu \quad ; \quad
\xi^4 = - {\cal R} \,  n(x) \left(1 - \epsilon \,
\frac{\sigma^2}{4 {\cal R}^2} \right) \; , 
\label{xix}
\ee
where
\be
n(x) = \frac{1}{1+ \epsilon \frac{\sigma^2}{4 {\cal R}^2}} \; ,
\label{n}
\ee
and
\be
\sigma^2 = \eta_{\mu \nu} \, x^\mu x^\nu \; ,
\ee
with $\eta_{\mu \nu} = \delta^{a}{}_{\mu} \, \delta^{b}{}_{\nu} \, \eta_{a b}$.
The $h^{a}{}_{\mu}$ introduced in (\ref{xix}) are the components of a tetrad field,
actually of the 1-form basis members $\omega^{a} = h^{a}{}_{\mu} dx^\mu = n \,
\delta^{a}{}_{\mu} dx^\mu$. The inverse transformations are
\be
x^\mu \equiv h_{a}{}^{\mu} \, \xi^{a} = n^{-1}(\xi) \, \delta_{a}{}^{\mu} \,
\xi^{a} \quad ; \quad
\epsilon \, \frac{\sigma^2}{4 {\cal R}^2} = 
\frac{1 + \xi^4/{\cal R}}{1 - \xi^4/{\cal R}} \; , 
\label{xxi}
\ee
where
\be
 n(\xi) = \frac{1}{2} \, \left( 1 - \frac{\xi^4}{{\cal R}} \right) \; .
\ee

In stereographic coordinates, the line element
$ds^2 = \eta_{AB} \, d\xi^A d\xi^B$ is found to be $ds^2 = g_{\mu \nu} \,d x^\mu dx^\nu$,
with
\be
g_{\mu \nu} = h^{a}{}_{\mu} \, h^{b}{}_{\nu} \, \eta_{a b} \equiv n^2(x) \, \eta_{\mu \nu}
\label{44}
\ee
the corresponding metric tensor. The de Sitter spaces, therefore, are conformally flat,
with the conformal factor given by $n^2$. We could have written simply $\xi^\mu = n \,
x^\mu$, but we are carefully using the Latin alphabet for the algebra (and flat space)
indices, and the Greek alphabet for the homogeneous space fields and cofields. As usual
with changes from flat tangent--space to spacetime, letters of the two kinds are
interchanged with the help of the tetrad field. This is true for all tensor indices.
Connections, which are vectors only in the last (1-form) index, will gain an extra
``vacuum" term~\cite{livro}.

The Christoffel symbol corresponding to the metric $g_{\mu \nu}$ is
\be
\Gamma^{\lambda}{}_{\mu \nu} = \left[ \delta^{\lambda}{}_{\mu}
\delta^{\sigma}{}_{\nu}  + \delta^{\lambda}{}_{\nu}
\delta^{\sigma}{}_{\mu} - \eta_{\mu \nu} \eta^{\lambda \sigma} \right]
\partial_\sigma (\ln n) \; .
\label{46}
\ee
The corresponding Riemann tensor components, 
\[
R^{\mu}{}_{\nu \rho \sigma} = \partial_\rho
\Gamma^{\mu}{}_{\nu \sigma} - \partial_\sigma \Gamma^{\mu}{}_{\nu \rho} +
\Gamma^{\mu}{}_{\epsilon \rho} \, \Gamma^{\epsilon}{}_{\nu \sigma} -
\Gamma^{\mu}{}_{\epsilon \sigma} \, \Gamma^{\epsilon}{}_{\nu \rho} \; ,
\]
are found to be
\be
R^{\mu}{}_{\nu \rho \sigma} = \epsilon \, \frac{1}{{\cal R}^2} \,
\left[\delta^{\mu}{}_{\rho} g_{\nu \sigma} - \delta^{\mu}{}_{\sigma} g_{\nu
\rho} \right] \; .
\label{47}
\ee
The Ricci tensor and the scalar curvature are, consequently
\be
R_{\mu \nu} = \epsilon \, \frac{3}{{\cal R}^2} \, g_{\mu \nu}
\label{48}
\ee
and
\be
R = \epsilon \, \frac{12}{{\cal R}^2} \;.
\label{49}
\ee
	
In terms of the coordinates $\{x^\mu\}$, the generators
(\ref{dsgene}) of the infinitesimal de Sitter transformations are given by
\ba
J_{a b} &\equiv& \delta_{a}{}^{\mu} \, \delta_{b}{}^{\nu} \,
\left( \eta_{\rho \mu} \, x^\rho \, P_\nu -
\eta_{\rho \nu} \, x^\rho \, P_\mu \right) \label{dslore} \\ {} \nonumber \\
J_{a 4} &\equiv& \epsilon \, \delta_{a}{}^{\mu} \, 
\left({\cal R} \, P_\mu + \frac{\epsilon}{4 {\cal R}} \, K_{\mu} \right) \; ,
\label{dstra}
\ea
where
\be
P_\mu = \frac{\partial}{\partial x^\mu} \quad {\rm and} \quad
K_\mu = \left(2 \, \eta_{\mu \lambda} \, x^\lambda \, x^\rho -
\sigma^2 \, \delta_{\mu}{}^{\rho} \right) P_\rho \; 
\ee
are respectively the generators of translations and special conformal transformations.
For $\epsilon = +1$, we get the generators of the de Sitter group $SO(4,1)$. For
$\epsilon = -1$, we get the generators of the de Sitter group $SO(3,2)$. 

To illustrate the whole process, we will follow the fate of a spinorial test particle in
the ensuing contraction procedures by writing the Dirac equation on these spaces.
Denoting by $\gamma^d$ the Dirac matrices, it is given by~\cite{dirac} 
\be 
i \, \hbar \, h_{d}{}^{\mu} \, \gamma^d \left[\partial_{\mu} -
\frac{i}{4} \omega^{ab}{}_{\mu} \, \sigma_{ab}\right] \psi(x) - m \, c \, \psi(x) = 0 \; ,
\ee
where $h_{d}{}^{\mu}$ = $n^{-1}(x) \delta_{d}{}^{\mu}$ is the inverse tetrad,
\be
\omega^{a}{}_{b \mu} = h^{a}{}_{\rho} (\partial_{\mu} h_{b}{}^{\rho} +
\Gamma^{\rho}{}_{\sigma \mu}h_{b}{}^{\sigma}) = ( h^{a}{}_{\mu}
h^{b}{}_{\sigma} -  h^{a}{}_{\sigma} h^{a}{}_{\mu}) \partial^\sigma (\ln n)
\ee
is the spin connection, and
\be
\sigma_{a b} = \frac{i}{2} \ \left[ \gamma_a , \gamma_b \right]
\ee
is the spin-1/2 representation of the Lorentz group. 
A direct calculation shows that
\be
n^{-3/2} \, \gamma^{\mu} \, \partial_{\mu} \left[ n^{3/2} \psi(x) \right] +
i \ n \ M \ \psi(x) = 0 \; ,
\label{direq}
\ee
with $M = m c/\hbar$, and where we have used the notation
\be
\gamma^\mu \equiv \delta^{\mu}{}_{d} \ \gamma^d  \; .
\ee

\section{WEAK COSMOLOGICAL--CONSTANT CONTRACTION}

The In\"on\"u--Wigner contraction of the de Sitter to the
Poincar\'e group is obtained by taking the limit ${\cal R} \rightarrow \infty$. In this
case, it is convenient to rewrite Eqs. (\ref{dslore}) and (\ref{dstra}) in the form
\ba
J_{a b} &\equiv& \delta_{a}{}^{\mu} \, \delta_{b}{}^{\nu} \,
L_{\mu \nu} \label{lore} \\ {} \nonumber \\
J_{a 4} &\equiv& {\cal R} \, \delta_{a}{}^{\mu} \, 
\Pi_\mu \; , \label{ds1tra}
\ea
where
\be
L_{\mu \nu} = \eta_{\rho \mu} \, x^\rho \, P_\nu -
\eta_{\rho \nu} \, x^\rho \, P_\mu
\label{lorentz}
\ee
are the generators of the Lorentz group, and
\be
\Pi_\mu = \epsilon \left( P_\mu + \frac{\epsilon}{4 {\cal R}^2} K_\mu \right)
\ee
the generators of the de Sitter translations. 
In terms of these generators, the commutation relation (\ref{dsal}) take the form 
\ba
\left[L_{\mu \nu}, L_{\lambda \rho}\right] &=& \eta_{\nu \lambda} \,
L_{\mu \rho} + \eta_{\mu \rho} \, L_{\nu \lambda} - \eta_{\nu \rho} \,
L_{\mu \lambda} - \eta_{\mu \lambda} \, L_{\nu \rho} \; , \label{ll} \\ {} \nonumber \\
\left[{\Pi}_{\mu}, L_{\lambda \rho}\right] &=& \eta_{\mu \lambda} \,
\Pi_{\rho} - \eta_{\mu \rho} \, \Pi_{\lambda} \; , \label{pl} \\ {} \nonumber \\
\left[{\Pi}_{\mu}, {\Pi}_{\lambda}\right] &=& - \epsilon \, {\cal R}^{-2}
\, L_{\mu \lambda} \; .
\label{pp}
\ea

Proceeding to the contraction limit ${\cal R} \rightarrow \infty$, we see that
\be
\lim_{{\cal R}\to\infty} \, L_{\mu \nu} = L_{\mu \nu} \quad \mbox{and} \quad
\lim_{{\cal R}\to\infty} \, \Pi_\mu = \epsilon \, P_\mu \; ,
\ee
and, in consequence, the de Sitter algebra contracts to the usual Poincar\'e algebra
\ba
\left[L_{\mu \nu}, L_{\lambda \rho}\right] &=& \eta_{\nu \lambda} \,
L_{\mu \rho} + \eta_{\mu \rho} \, L_{\nu \lambda} - \eta_{\nu \rho} \,
L_{\mu \lambda} - \eta_{\mu \lambda} \, L_{\nu \rho} \; , \\ {} \nonumber \\
\left[P_{\mu}, L_{\lambda \rho}\right] &=& \eta_{\mu \lambda} \,
P_{\rho} - \eta_{\mu \rho} \, P_{\lambda} \; , \\ {} \nonumber \\
\left[P_{\mu}, P_{\lambda}\right] &=& 0 \; .
\ea
We see also that 
\[
\lim_{{\cal R}\to\infty} \, g_{\mu \nu} = \eta_{\mu \nu} \; ,
\]
which shows that this limit leads exactly to the Minkowski geometry, a geometry
gravitationally related to a zero cosmological constant. Correspondingly, the Dirac
equation (\ref{direq}) acquires the expected form,
\be
\gamma^{\mu} \partial_{\mu}\psi(x) + i \ M \ \psi(x) = 0.
\ee

\section{STRONG COSMOLOGICAL--CONSTANT CONTRACTION}
	
Let us consider now the opposite limit, that is, ${\cal R}
\rightarrow 0$. In this case, we rewrite Eqs. (\ref{dslore}) and
(\ref{dstra}) in the form
\ba
J_{a b} &\equiv& \delta_{a}{}^{\mu} \, \delta_{b}{}^{\nu} L_{\mu \nu} \\ {} \nonumber \\
J_{a 4} &\equiv& {\cal R}^{-1} \, \delta_{a}{}^{\mu} \, \kappa_{\mu}  \; ,
\ea
where $L_{\mu \nu}$ are the generators of the Lorentz group, and
\be
\kappa_{\mu} = \frac{1}{4} \, K_\mu + \epsilon \, {\cal R}^2 \, P_{\mu} \; .
\label{pmuz}
\ee
In terms of these generators, the commutation relation (\ref{dsal}) becomes
\ba
\left[L_{\mu \nu}, L_{\lambda \rho}\right] &=& \eta_{\nu \lambda} \,
L_{\mu \rho} + \eta_{\mu \rho} \, L_{\nu \lambda} - \eta_{\nu \rho} \,
L_{\mu \lambda} - \eta_{\mu \lambda} \, L_{\nu \rho} \; , \label{llz} \\ {} \nonumber \\
\left[\kappa_{\mu},L_{\lambda \rho}\right] &=& \eta_{\mu \lambda} \,
\kappa_{\rho} - \eta_{\mu \rho} \, \kappa_{\lambda} \; , \label{plz} \\ {} \nonumber \\
\left[\kappa_{\mu}, \kappa_{\lambda}\right] &=& - \epsilon \, {\cal R}^{2}
\, L_{\mu \lambda} \; .
\label{ppz}
\ea

In the contraction limit ${\cal R} \rightarrow 0$, one can see that
\be
\lim_{{\cal R}\to 0} \,L_{\mu \nu} = L_{\mu \nu} \quad ; \quad
\lim_{{\cal R}\to 0} \, \kappa_\mu = \frac{1}{4} \, K_\mu \; ,
\ee
and consequently the de Sitter algebra contracts to 
\ba
\left[L_{\mu \nu}, L_{\lambda \rho}\right] &=& \eta_{\nu \lambda} \,
L_{\mu \rho} + \eta_{\mu \rho} \, L_{\nu \lambda} - \eta_{\nu \rho} \,
L_{\mu \lambda} - \eta_{\mu \lambda} \, L_{\nu \rho} \; , \label{llx} \\ {} \nonumber \\
\left[ K_{\mu}, L_{\lambda \rho}\right]&=&\eta_{\mu \lambda}  K_{\rho} -
\eta_{\mu \rho}  K_{\lambda} \; , \label{klx} \\ {} \nonumber \\
\left[ K_{\mu},  K_{\lambda}\right]&=&0 \; .
\label{kk}
\ea

The Lie group corresponding to this algebra, denoted by $Q$ and formed by a semi--direct
product of Lorentz and special conformal transformations, is completely different from
$P$ but presents the same Lie algebra as the Poincar\'e group. It will rule the local
kinematics of high $\Lambda$ spaces. We see also that, for
${\cal R} \rightarrow 0$, the asymptotic behaviour of the conformal factor $n(x)$ is
\[
n(x) \sim \epsilon\, \frac{4 {\cal R}^2}{\sigma^2} \; .
\]
The metric tensor, thus, becomes singular,
\[
\lim_{{\cal R}\to 0} \, g_{\mu \nu} = 0 \; ,
\]
and the Riemann and Ricci curvature tensors vanish, as can be seen from Eqs. (\ref{47})
and (\ref{48}), respectively. However, the scalar curvature $R$ becomes infinity, which is
in accordance with the fact that in this limit the cosmological constant goes to
infinity. We can conclude, therefore, that the contraction limit ${\cal R} \rightarrow
0$ leads both de Sitter spaces to a spacetime, denoted by $N$, whose geometry is
gravitationally related to an infinite cosmological constant. It is a 4-dimensional
cone--space in which $ds = 0$, and whose group of motion is $Q$. Analogously to the
Minkowski case, $N$ is also a homogeneous space, but now under the kinematical group $Q$,
that is, $N = Q/L$. In other words, the point--set of $N$ is the point--set of the
special conformal transformations. Furthermore, the manifold of $Q$ is a principal bundle
${\cal P}(Q/L,L)$, with $Q/L \equiv N$ as base space and $L$ as the typical fiber. 

The kinematical group $Q$, like the Poincar\'e group, has the Lorentz group $L$ as the
subgroup accounting for both the isotropy and the equivalence of inertial frames in this
space. However, the special conformal transformations introduce a new kind of
homogeneity. Instead of ordinary translations, all the points of $N$ are equivalent
through special conformal transformations.

This 4-dimensional cone--space should not, of course, be confused with the
3-dimen\-sion\-al light--cone of special relativity. Nevertheless, due to the
conformal factor $n(x)$, the mass term in (\ref{direq}) vanishes and the Dirac equation
takes the Weyl form
\be
\sigma^{3} \, \gamma^{\mu} \, \partial_{\mu} \left[ \sigma^{-3}\psi(x) \right] = 0 \; ,
\ee 
which can be reduced to a two--component equation. It is not surprising that, in the
presence of the newly--acquired conformal symmetry, the mass come to pass out of sight.

\section{FINAL REMARKS}

By the process of In\"on\"u--Wigner group contraction with ${\cal R} \rightarrow \infty$,
both de Sitter groups are reduced to the Poincar\'e group $P$, and both de Sitter
spacetimes are reduced to the Minkowski space $M$. As the de Sitter scalar curvature goes
to zero in this limit, we can say that $M$ is a spacetime gravitationally related to a
vanishing cosmological constant. On the other hand, in a similar fashion but taking the
limit ${\cal R} \rightarrow 0$, both de Sitter groups are contracted to the group $Q$,
formed by a semi--direct product between Lorentz and special conformal transformation
groups, and both de Sitter spaces are reduced to the cone--space $N$, which is a space
with vanishing Riemann and Ricci curvature tensors. As the scalar curvature of the
de Sitter space goes to infinity in this limit, we can say that $N$ is
a spacetime gravitationally related to an infinite cosmological constant. If
the fundamental spacetime symmetry of the laws of Physics is that given by the de Sitter
instead of the Poincar\'e group, the $P$-symmetry of the weak cosmological--constant
limit and the $Q$-symmetry of the strong cosmological--constant limit can be considered
as limiting cases of the fundamental symmetry.

Minkowski and the cone--space can be considered as {\it dual} to each other, in the
sense that their geometries are determined respectively by a vanishing and an infinite
cosmological constants. The same can be said of their kinematical group of motions: $P$
is associated to a vanishing cosmological constant and $Q$ to an infinite cosmological
constant. The {\em dual} transformation connecting these two geometries is the spacetime
inversion
\[
x^\mu \rightarrow - \frac{x^\mu}{\sigma^2} \; .
\]
Under such a transformation, the Poincar\'e group $P$ is transformed into the group $Q$,
and the Minkowski space $M$ becomes the cone--space $N$. The points at infinity of $M$
are concentrated in the vertex of the cone--space $N$, and those on the light--cone of
$M$ becomes the infinity of $N$. It is interesting to notice that, despite presenting an
infinite
scalar curvature, the concepts of space isotropy and equivalence between inertial frames
in the cone--space $N$ are those of special relativity. The difference lies in the
concept of uniformity as it is the special conformal transformations, and not ordinary
translations, which act transitively on $N$.

Besides presenting an intrinsic mathematical and physical interest, in the light of the
recent supernovae results~\cite{novae} favoring possibly quite large values for the
cosmological constant, the above results may acquire a further relevance to Cosmology,
with applications to inflationary models as well.

\section*{ACKNOWLEDGMENTS}

The authors would like to thank FAPESP--Brazil and CNPq--Brazil for partial financial
support.

\end{document}